\documentstyle[preprint,aps,prl]{revtex}
\begin{document}
\draft
\title{ Magnetic susceptibility of insulators from first 
principles} 
\author{Francesco Mauri and Steven G. Louie}
\address{Department of Physics, University of California at Berkeley,
Berkeley, CA 94720, USA \\
and Materials Science Division, Lawrence Berkeley Laboratory, Berkeley, 
CA 94720, USA}
\maketitle 

\begin{abstract} 
We present an {\it ab initio} approach for the computation of the magnetic 
susceptibility $\chi$ of  insulators. 
The approach is applied to compute $\chi$ in diamond and in solid neon using 
density functional theory in the local density approximation, obtaining
good agreement with experimental data. 
In solid neon, we predict an observable dependence of $\chi$ upon pressure. 
\end{abstract}


\narrowtext

The response of an extended system to a uniform external magnetic field is 
a fundamental property. This response can be used as a sensitive probe to
the structural and electronic properties of materials, 
such as in the case of nuclear magnetic resonance spectroscopy.
However, to our knowledge, 
the orbital magnetic susceptibility $\chi$
of real solids has not been computed from first principles. 
In this work we discuss an {\it ab initio} approach
for the evaluation of $\chi$ in insulators within
density functional theory (DFT).
We applied our formalism to diamond and solid neon using the local 
density approximation (LDA)
for the exchange and correlation energy.
The agreement of our results with experimental data indicates that
DFT-LDA describes correctly the magnetic response of these systems.

The susceptibility $\chi$ has been evaluated in cubic semiconductors using
empirical methods\cite{SW74}.
Exact expressions for $\chi$ of a periodic solid in terms of 
Bloch eigenstates and eigenvalues, have been derived already in 
the sixties\cite{B61,R61,HLSS64}. However these approaches are rather 
involved and have not been applied to real materials.
A more compact expression for $\chi$ was recently given in Ref. \cite{V91}, 
where it is applied to a model 2-dimensional system.
Our approach for the computation of $\chi$ in real systems is related to 
the one of Ref. \cite{V91}. 

The paper is organized as follows.
First we present the formalism for a generic single particle Hamiltonian.
Then we justify the use DFT in the LDA in the computation of $\chi$, and 
we discuss the accuracy and the limits of the additional use of the 
pseudopotential approximation.
Finally, we apply our formalism to diamond and solid neon, studying the 
behavior of $\chi$ as a function of the lattice constant.

The magnetic susceptibility is defined as the second derivative of the total
energy per unit volume $E$ with respect to the macroscopic magnetic field 
${\bf B}$, i.e.:
\begin{equation}
\chi_{ij}= -{d^2 E\over d B_i d B_j},
\end{equation}
where $i$ and $j$ are the Cartesian indexes. 
To simplify the notation in the following discussion, 
we consider a cubic system for which
$\chi_{ij}=\delta_{ij}\chi$.

Perturbation theory can be used to compute $\chi$.
This is straightforward for a finite system.
However, in an extended solid, the expectation values of the perturbative 
Hamiltonian on delocalized eigenstates are not well-defined quantities 
for an uniform field.
To avoid this problem we consider the response of the system to a magnetic 
field with a finite wavelength ${\bf q}=(q,0,0)$, 
i.e. ${\bf B}(x) =b(0,0,\sqrt{2}\cos (q x))={\bf \nabla}\times {\bf A} $
with ${\bf A}(x)=b(0,\sqrt{2}{\sin (q x)/ q},0)$.
Defining
\begin{equation}
\chi(q)=-{d^2 E\over d b^2},
\end{equation}
in the limit of $q\rightarrow 0$, we obtain the macroscopic 
susceptibility $\chi$\cite{HLSS64}.

Let us first consider a system described by a single particle Hamiltonian.
If the coupling between ${\bf B}$ and the spin of the electron can be neglected, 
the perturbation to the Hamiltonian can be written as 
$\Delta H=H^{(1)}+H^{(2)}$ with
\begin{eqnarray}
\label{defh}
H^{(1)}&=&{1\over c}{\bf p}\cdot{\bf A}={\sqrt{2}\over c}{\sin (q x)\over q} 
p_y b, \nonumber \\
H^{(2)}&=&{1\over 2c^2}A^2={1\over c^2}{\sin^2 (q x)\over q^2 }b ^2,
\end{eqnarray}
where atomic unit are used, ${\bf p}$ is the momentum operator, 
and $c$ is the speed of light.

For a periodic insulator we have:
\begin{eqnarray}
\label{per1}
\chi(q)b^2&=&-4{\Omega\over c^2}
\int {d^3{\bf k}\over 8\pi^3}
\int {d^3{\bf k}'\over 8\pi^3}
\sum_{i\in {\cal O},j \in {\cal E}}
{ 
  |\langle \psi_{{\bf k},i}|H^{(1)}|\psi_{{\bf k}',j}\rangle|^2
                         \over 
\epsilon_{{\bf k},i}-\epsilon_{{\bf k}',j} } \nonumber \\
& &-{4\over c^2}   \int {d^3{\bf k} \over 8\pi^3} \sum_{i\in {\cal O} }
\langle \psi_{{\bf k},i}|H^{(2)}|\psi_{{\bf k},i}\rangle,
\end{eqnarray}
where $\psi_{{\bf k},i}$ and $\epsilon_{{\bf k},i}$ are the Bloch
eigenstates and eigenvalues of the unperturbed Hamiltonian, $\Omega$ is the 
volume of the unit cell, 
$\cal O$ and $\cal E$ are the sets of   
occupied and empty bands,
and a factor of 2 for spin degeneracy is included.
By inserting Eq.~(\ref{defh}) in Eq.~(\ref{per1}), we get:
\begin{eqnarray}
\label{per2}
\chi(q)&=&-{2\over c^2q^2}\int {d^3{\bf k}\over 8\pi^3} 
[g({\bf k}+{\bf q},{\bf k})+g({\bf k}-{\bf q},{\bf k})] \nonumber \\
& & -{N\over\Omega c^2q^2} 
\end{eqnarray}
where $N$ is the number of electrons per unit cell,
\begin{equation}
g({\bf k}',{\bf k})=\sum_{i\in {\cal O},j
\in {\cal E}}{|\langle u_{{\bf k}',i}
|-i\nabla_y+{k'_y+k_y\over 2}|u_{{\bf k},j}\rangle|^2\over
\epsilon_{{\bf k}',i}-\epsilon_{{\bf k},j}},\label{defg}
\end{equation}
and $|u_{{\bf k},i}\rangle$ is the periodic part of the Bloch eigenstate
(normalized in the unit cell).
For $q\rightarrow 0$, the two terms on the right-hand-side (rhs) 
of Eq.~(\ref{per2}) 
individually diverge, but $\chi(q)$ remains finite.
To show this, we use the f-sum rule:
\begin{equation}
f_{s}={N\over\Omega}=-4 \int {d^3{\bf k}\over 8\pi^3} g({\bf k},{\bf k}).
\label{fsumrule}
\end{equation}
By inserting Eq.~(\ref{fsumrule}) in Eq.~(\ref{per2}) we obtain:
\begin{equation}
\chi(q)=-{2\over c^2}\int {d^3{\bf k}\over 8\pi^3}{g({\bf k}+{\bf q},{\bf k})-2
g({\bf k},{\bf k})+g({\bf k}-{\bf q},{\bf k})\over q^2}.
\label{ximq}
\end{equation}
Then
\begin{equation}
\chi=\lim_{q\rightarrow 0}\chi(q)=
-{2\over c^2}\int {d^3{\bf k}\over 8\pi^3} 
{d^2\over dk_x^2}g({\bf k},{\bf k}')|_{{\bf k}'={\bf k}}.\label{xim}
\end{equation}
Similar conclusions have been obtained in Ref. \cite{V91}.

In our numerical evaluation of the macroscopic $\chi$ we use 
Eq.~(\ref{ximq}) with a small but finite $q$.
Note that Eq.~(\ref{per2}) is not suitable to this approach.
Indeed, in a practical application, both the integral in $\bf k$ space and the 
sum over all empty bands are replaced by finite sums.
Under these conditions the f-sum rule, Eq.~(\ref{fsumrule}), 
is no longer exactly satisfied.
Then for $q\rightarrow 0$ the rhs of Eq.~(\ref{per2})  will diverge as 
$\Delta f_{s}c^{-2}q^{-2}$, where  $\Delta f_s$ is the the error in the 
f-sum rule. This numerical instability does not occur 
in Eq.~(\ref{ximq}) where every term
is treated consistently.

We computed $\chi$ using DFT-LDA, i.e. we   
neglected any explicit dependence of the exchange-correlation 
functional on the current density.
Ref.~\cite{cdft} proposes an approximate functional for the 
exchange correlation energy $E_{xc}$
which depends also on the current.
The current term in $E_{xc}$ influences the magnetic response 
in systems with a small electronic density.
It is negligible in our case, since 
it yields a correction to $\chi$ smaller than 2$\%$
at the electronic densities typical of the systems we are studying\cite{cdft,VRG88}.
We also do not consider magnetic local field effects, which are negligible 
in non-magnetic materials\cite{CGSunit}.
Finally, we note that the DFT Hamiltonian depends in a self-consistent way upon the
electronic charge density. Thus, in general, to compute the second order variation
in the total energy with respect to an external perturbation, 
one should take into account
the linear variation of the Hamiltonian induced by the linear 
variation of the charge $\delta \rho$ (see e.g. Ref.~\cite{BGT}). 
However, if the perturbation is a magnetic field, $\delta \rho$ is zero
by time reversal symmetry. Thus Eq.~(\ref{ximq}) is correct 
within DFT.

In our present practical calculation we used the 
pseudopotential approach, in which only the valence electrons are 
considered.
To discuss the validity of the pseudopotential 
approximation in the computation of $\chi$,
we divide the set of occupied bands ${\cal O}$ into the sets of core 
bands ${\cal C}$ and valence bands ${\cal V}$.
Then we have:
\begin{equation}
\chi=  \chi_{{\cal C},{\cal E}}+
\chi_{{\cal V},{\cal E}}
=\chi_{{ C}}-\chi_{{\cal C},{\cal V}}+\chi_{{\cal V},{\cal E}}.\label{ximcv}
\end{equation}
Here $\chi_{{\cal C},{\cal E}}$ is given by  
Eqs.~(\ref{defg}) and (\ref{xim}) with the sum over the $i$ and $j$ indexes 
in Eq.~(\ref{defg}) running over the 
the sets of core states, ${\cal C}$, and of empty states, ${\cal E}$, respectively. 
The other $\chi$ with two indices
are define in a similar way. $\chi_{{ C}}$
is the magnetic susceptibility of the core electrons, which is not sensitive on the 
chemical environment and thus can be computed considering the isolated atoms, i.e.:
\begin{equation}
\chi_{{ C}}=\chi_{{\cal C},{\cal E}}+\chi_{{\cal C},{\cal V}}\simeq
-{1\over\Omega c^2}\sum_I\sum_{i\in {\cal C}}\langle \Psi_i^I|x^2|\Psi_i^I\rangle,
\label{ximc}
\end{equation}
where we sum over the atoms in the unit cell, and $\Psi_i^I$
are the core atomic wavefunctions of the atom $I$.
Among the three terms in the rhs of Eq.~(\ref{ximcv}),
$\chi_{{\cal V}, {\cal E}}$ is the only one accessible in a pseudopotential 
calculation; 
$\chi_{{ C}}$ can be computed using an atomic code, but the evaluation of 
$\chi_{{\cal C},{\cal V}}$ requires the knowledge of 
both core and valence wavefunctions.
Since  $\chi_{{\cal C},{\cal V}}$ and
$\chi_{{ C}}$ are expected to be of the same order of magnitude,
the pseudopotential approximation
introduces an error of the order of $\chi_{{ C}}$ by neglecting $\chi_{{\cal C},{\cal V}}$.
This error is reasonably small only 
for elements in the first and 
second row of the Periodic Table, for which $\chi_{{ C}}\ll \chi$.
For application of the present theory to heavier elements, all-electron 
calculations are needed.
Finally, in our pseudopotential calculation, we replaced the operator
$-i{\bf \nabla} +{\bf k}$ in Eq.~(\ref{defg}) with the velocity operator 
${\bf v}^p_{\bf k} = (d/d{\bf k})H^p_{{\bf k}}$ where $H^p_{{\bf k}}$ is the
pseudo-Hamiltonian\cite{footnotev}.

We computed $\chi$ for isolated carbon (C) and neon (Ne) atoms, 
for solid Ne in the fcc
structure, and for solid C in the diamond structure.
In the atomic phases we used the all-electron ground state wavefunctions
to compute $\chi_{C}$ using
Eq.~(\ref{ximc}). In Ne we also computed the atomic $\chi$  
using Eq.~(\ref{ximc}) with the sum over the index $i$ running over
all occupied states.
In the solid phases, we evaluated $\chi_{{\cal V },{\cal E}}$ using
Eq.~(\ref{ximq}) with a $q={.03\pi/ a}$, where $a$ is the lattice constant of the
cubic cell. 
The pseudopotentials were generated using the prescription of 
Ref.~\cite{TM91}. 
In Ne we expanded the wavefunctions on a 
plane-wave basis set with a 120 Ry cutoff.
We sampled the $\bf k$ space integrals with 10 special $\bf k$-points
in the irreducible Brillouin zone, and considered 400 empty states. 
In diamond we used a 60 Ry cutoff, 60 special $k$-points, and 300 empty states. 
We verified that with the above parameters the convergence error in the value 
of $\chi$ is less than 0.2$\%$.

The results for Ne are shown in Table~\ref{tableNe}.
The atomic calculation is in good agreement with 
the experimental data.
For the solid fcc
phase we report $\chi_{{\cal V },{\cal E}}$ 
as a function of the lattice constant $a$.
We note that $\chi_{{\cal V },{\cal E}}$ reaches a plateau for $a\sim a^e_0$,
where $a^e_0$ is the experimental equilibrium lattice constant. 
This indicates that for $a\ge a^e_0$
the interaction among Ne atoms is negligible.
Moreover $\chi_{{\cal V },{\cal E}}$ at $a=a_0$ is very close to the value of $\chi$ 
computed for the isolated atom.
This establishes, in the atomic limit, the correctness of our approach and the 
accuracy of the pseudopotential approximation. 
As $a$ decreases, $-\chi_{{\cal V },{\cal E}}$ decreases.
This can be understood by noting that for an isolated closed shell atom 
only a negative diamagnetic term contributes to $\chi$, since the 
unperturbed Hamiltonian is spherically symmetric.
As the Ne atoms get closer, spherical symmetry is broken and a 
positive paramagnetic term also contributes to $\chi$.
For the sake of comparison with future experiments, 
we also report the theoretical pressure P as a function of $a$.
Solid Ne at zero P is bonded by a weak van der Waals interaction,
which is incorrectly biven by LDA\cite{PB95}. 
Thus for the larger $a$ we do not expect to obtain
accurate values for P.
However we expect LDA to describe correctly the repulsive interaction 
between Ne atoms, which dominates P at smaller $a$.
Note that at P=50 GPa, $-\chi_{{\cal V },{\cal E}}$ is decreased
by 16$\%$ with respect to its atomic value.

The results for $C$ are shown in Table~\ref{tableC}.
Since C is not a closed shell atom, in the atomic case only 
$\chi_{C}$ is reported.
In the diamond phase we  report $\chi_{{\cal V },{\cal E}}$
as a function of the lattice constant $a$.
The computed pressure obtained from the LDA-DFT total energies is
also shown.
In the range of experimentally accessible pressures 
$\chi_{{\cal V },{\cal E}}$ shows a 
negligible dependence upon $a$.
Both the values of $\chi_{{\cal V },{\cal E}}$ at the experimental ($a_0^e$) and at the 
theoretical ($a_0^t$)
equilibrium lattice constant are in very good agreement with the experimental 
data. 

In conclusion we have presented a method to compute the magnetic response
of real solids from first principles. We have shown that DFT-LDA reproduces the
magnetic susceptibility $\chi$ of diamond. In diamond 
$\chi$ is found to be insensitive to the applied pressure 
whereas we predict an observable pressure dependence
of $\chi$ in solid Ne.

We thank V. Crespi and O. Zakharov for a critical reading of the manuscript.
This work was supported by the National Science
Foundation under Grant No. DMR-9520554, by the Office of Energy Research, Office of 
Basic Energy Sciences, Materials Sciences Division of the U.S. Department of Energy
under Contract No. DE-AC03-76SF00098,
and by the Miller Institute for Basic Research in Science.
Computer time was provided by the NSF at the Pittsburg Supercomputing Center.

\begin{table}

\caption{ 
\label{tableNe}
Magnetic susceptibility of atomic and solid fcc Ne in units of
10$^{-6}$cm$^3$/mole. 
In the solid we
considered different values
of the lattice constant $a$.
We indicate with $a_0^e$
the experimental equilibrium lattice constant.
The theoretical pressure P is also reported.}

\begin{tabular}{lcccc}
  &$-\chi$&$-\chi_{C}$&$-\chi_{{\cal V},{\cal E}}$&P (GPa) \\
\tableline
Atom (experiment)      & 7.2   &       &      &      \\
Atom (theory)          & 7.80  & .05   &      &      \\
Solid $a=$8.37au$=a_0^e$ &   &       & 7.79 & -2   \\
Solid $a=$7.87au     &       &       & 7.76 & -2   \\
Solid $a=$7.37au     &       &       & 7.64 & -1   \\
Solid $a=$6.87au     &       &       & 7.41 &  4   \\
Solid $a=$6.37au     &       &       & 7.14 &  15  \\
Solid $a=$5.87au     &       &       & 6.66 &  50  \\
Solid $a=$5.37au     &       &       & 6.04 &  151\\
\end{tabular}
\end{table}

\begin{table}
\caption{ 
\label{tableC}
Magnetic susceptibility of atomic C 
and of diamond 
in units of 10$^{-6}$cm$^3$/mole of C$_2$. For the solid we 
considered different values
of the lattice constant $a$. We indicate with 
$a_0^e$  and $a_0^t$ the experimental and theoretical 
equilibrium lattice constants, respectively.
The theoretical pressure P is also reported.}

\begin{tabular}{lcccc}
  & $-\chi$ & $-\chi_{C}$ & $-\chi_{{\cal V },{\cal E}}$& P (GPa) \\
\tableline
Solid (experiment) & 11.8 &       &       &   \\
Atom (theory)      &       & 0.32 &       &   \\
Solid $a=$6.75au$=a_0^e$ &       &       & 11.17&  -17 \\
Solid $a=$6.66au$=a_0^t$ &       &       & 11.23&  0 \\
Solid $a=$6.55au    &       &       & 11.26&  25 \\
Solid $a=$6.35au    &       &       & 11.23&  85 \\
Solid $a=$6.15au    &       &       & 11.16&  168 \\
Solid $a=$5.95au    &       &       & 11.09&  283 \\
\end{tabular}
\end{table}

\end{document}